\documentclass[aps,prb,twocolumn,superscriptaddress,english]{revtex4}
\usepackage{amssymb}
\usepackage{graphicx}
\usepackage{amsmath}
\usepackage{soul}
\usepackage{amsthm}
\usepackage{amsfonts}
\usepackage[T1]{fontenc}
\usepackage[latin9]{inputenc}
\usepackage{array}
\usepackage{multirow}
\usepackage{color}
\usepackage{esint}
\usepackage{bm}
\usepackage{color}
\usepackage{bbm}
\usepackage{hyperref}
\usepackage{babel}
\usepackage{titlesec}

\begin{document}

\global\long\def\id{\mathbbm{1}}
\global\long\def\ui{\mathbbm{i}}
\global\long\def\ud{\mathrm{d}}

\title{Level spacing distribution of localized phases induced by quasiperiodic potentials}
\author{Chao Yang}
\affiliation{Shenzhen Institute for Quantum Science and Engineering,
Southern University of Science and Technology, Shenzhen 518055, China}
\affiliation{International Quantum Academy, Shenzhen 518048, China}
\affiliation{Guangdong Provincial Key Laboratory of Quantum Science and Engineering, Southern University of Science and Technology, Shenzhen 518055, China}
\author{Yucheng Wang}
\thanks{Corresponding author: wangyc3@sustech.edu.cn}
\affiliation{Shenzhen Institute for Quantum Science and Engineering,
	Southern University of Science and Technology, Shenzhen 518055, China}
\affiliation{International Quantum Academy, Shenzhen 518048, China}
\affiliation{Guangdong Provincial Key Laboratory of Quantum Science and Engineering, Southern University of Science and Technology, Shenzhen 518055, China}
\begin{abstract}
Level statistics is an important quantity for exploring and understanding localized physics. The level spacing distribution (LSD) of the disordered localized phase follows Poisson statistics, and many studies naturally apply it to the quasiperiodic localized phase. Here we analytically obtain the LSD of the quasiperiodic localized phase, and find that it deviates from Poisson statistics. Moreover, based on this level statistics, we derive the ratio of adjacent gaps and find that for a single sample, it is a $\delta$ function, which is in excellent agreement with numerical studies. Additionally, unlike disordered systems, in quasiperiodic systems, there are variations in the LSD across different regions of the spectrum, and the presence of spectral correlations results in non-equivalence between increasing the size and increasing the sample. Our findings carry significant implications for the reevaluation of level statistics in quasiperiodic systems and a profound understanding of the distinct effects of quasiperiodic potential-induced and disorder-induced localization.
\end{abstract}
\maketitle

\section{Introduction}
Quantum localization has consistently been a significant research area in condensed matter physics.
This phenomenon is widely present in disordered systems, caused by interference from multiply scattered waves due to system disorder, resulting in the exponential decay of the wave function and the suppression of transport~\cite{Anderson1958,RMP1,RMP2,Kramer1993}.
In addition to random disorder, quasiperiodic potentials also induce localization, and in recent years, they have garnered widespread interest in both theoretical~\cite{Soukoulis1981,DasSarma1988,Biddle2009,XLi2017,HYao2019,Ganeshan2015,Wang1,XCZhou,Wang2022,XDeng2019,Ribeiro} and experimental~\cite{Roati2008,Bloch4,An2018,JiasT,TXiao2021,Weld,HengFan} aspects, playing a crucial role in enhancing our understanding of critical phases~\cite{TXiao2021,Weld,HengFan,WangYC2021,Kohmoto1990}, rich transport behaviors~\cite{LandiRMP,Saha,Dwiputra,Lacerda,Jeffrey}, many-body localization (MBL)~\cite{IBlochRMP,BAA,Schreiber2015}, low dimensional Anderson transition (AT) and mobility edges~\cite{Soukoulis1981,DasSarma1988,Biddle2009,XLi2017,HYao2019,Ganeshan2015,Wang1,XCZhou,Wang2022,XDeng2019,Ribeiro,Roati2008,Bloch4,An2018,JiasT}. Furthermore, moir\'{e} materials have attracted considerable attention recently. Quasiperiodic modulations can manifest naturally in moir\'{e} materials~\cite{SJAhn,BHuang2019,DMao2021,Castro2022,AUri2023}. Specifically, by mapping strained moir\'{e} systems in a uniform magnetic field to a one-dimensional (1D) quasiperiodic system~\cite{Jeffrey,LiangFu}, we can gain insights into some intriguing properties of moir\'{e} materials.


The level spacing distribution (LSD) of localized phases is completely distinct from that of extended phases, allowing us to use LSD to differentiate between extended and localized phases~\cite{Shklovskii1993,Mirlin2000}. For disordered systems, the energy levels of the localized phase are uncorrelated, with no level repulsion, and their distribution follows Poisson statistics~\cite{Shklovskii1993,Mirlin2000,Molcanov1981}. Extending the statistical patterns of level spacing for disorder-induced localized phases to quasiperiodic-induced localized phases seems natural. Additionally, the average of the adjacent gap ratio $\langle r\rangle$ is close to 0.387~\cite{BHuang2019,XLi2016,YLiu2020,YWang2023,Xianlong,Ray2016,JHPixley,Schiffer2021,Ray2018,Khemani2017,Roushan2017,XiaopengLi,Modak2015}, which is in complete agreement with the results predicted by Poisson statistics. 
Therefore, the LSD of quasiperiodic localization systems, including the quasiperiodic localization in moir\'{e} systems, are widely accepted to follow Poisson statistics in both single-particle~\cite{BHuang2019,XLi2016,YLiu2020,YWang2023,Xianlong,Ray2016,JHPixley,Schiffer2021,Ray2018,Machida1986,SNEvangelou,Roy2019,Takada2004,YWang2016} and many-body systems~\cite{Khemani2017,Roushan2017,XiaopengLi,Modak2015,Vu2022}.
However, recent mathematical proof has shown that the distribution of eigenvalues in quasiperiodic and disordered localized phases exhibits significant differences~\cite{Jitomirskaya}, implying that the patterns of energy level spacings for the two cases may also differ. Therefore, it is necessary to reexamine the distribution of level spacings in quasiperiodic systems. This is helpful for understanding various properties of quasiperiodic systems, including moir\'{e} quasicrystals, as well as distinguishing between quasiperiodic localization and disordered localization.

In this letter, we take the Aubry-Andr\'{e} (AA) model as an example to investigate the LSD of the quasiperiodic localized phase. We first compare the energy level distribution, LSD $P(\delta E)$ and the distribution of the adjacent gap ratio $P(r)$ for Anderson localization (AL) phase induced by quasiperiodic potentials with those induced by disorder. Then, we calculate the number variance of different regions of energy spectrum in the AA model~\cite{AA} and compare it with the number variance of the levels in the AL phase induced by disorder. Such comparisons demonstrate the differences in the level distributions between quasiperiodic potentials and disorder-induced AL phases, intuitively and quantitatively showing that the levels of the AL phase in quasiperiodic systems are repulsive, meaning they are correlated, and therefore their spacing distributions are not Poisson. Finally, we analytically derive $P(\delta E)$, $P(r)$ and $\langle r\rangle$ for the AA model's AL phase. 

\begin{figure*}
	\centering
	\includegraphics[width=0.95\textwidth]{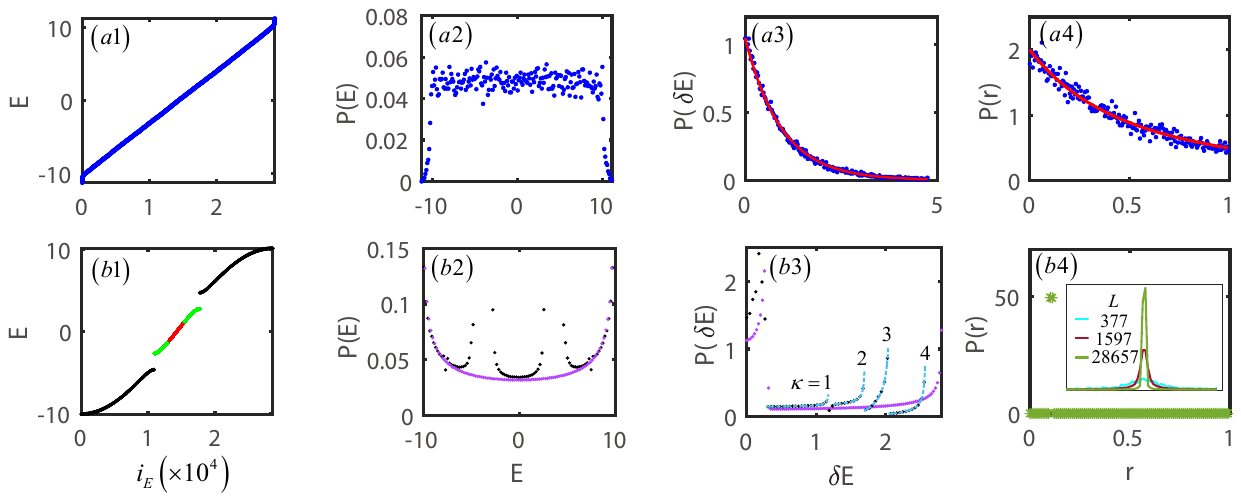}
	\caption{\label{01} (a1-a4) are about disorder-induced localization with $W/J=10$, and (b1-b4) are about localization induced by quasiperiodic potentials with $\alpha=F_{22}/F_{23}$, $\theta=0.3\pi$ and $V/J=10$. The size of both systems is $L=F_{23}=28657$. (a1) and (b1) Eigenenergies in ascending order, and the index of energy mode $i_E$ runs from $1$ to $L$. The green/red/green curves in (b1) correspond to the lowest/middle/highest $1/3$ energies of the middle region. (a2) and (b2) The distribution of energies (density of states). The purple data points in (b2) correspond to $J=0$. (a3) and (b3) Level spacing statistics with $\delta E=\Delta E/\langle\Delta E\rangle$ and $\langle\Delta E\rangle$ is the mean level spacing. The red fitting curve in (a3) shows $P(\delta E)=1.047e^{-1.065\delta E}$. The purple and light blue dashed lines in (b3) correspond to the fitting results using Eq.~(\ref{limt}) (corresponding to $J=0$) and Eq.~(\ref{pdeltaE}) (the fitting parameters corresponding to the branches $\kappa = 1, 2, 3, 4$ are $A_{\kappa} = 0.092, 0.31, 1.701, 0.595$, $b_{\kappa}L = 1.182, 1.7, 2.071, 2.571$, and $C_{\kappa} = 0.125, 0.099, -0.357, -0.079$), respectively. (a4) and (b4) The distributions of $P(r)$. The red curve in (a4) satisfies $P(r)=2/(1+r)^2$. Inset of (b4): The behavior $P(r)$ with different sizes. }
\end{figure*}

\section{Model and results}
The AA model is the simplest nontrivial example with a 1D quasiperiodic potential,
described by
\begin{equation}\label{AAH}
	H=J\sum_{j}(c_{j+1}^{\dagger}c_{j}+c_{j}^{\dagger}c_{j+1})+\sum_{j}V_jc_{j}^{\dagger}c_{j},
\end{equation}
where $c_j$ ($c^{\dagger}_j$) denotes the annihilation (creation) operator at site $j$, $J$ is the nearest-neighbor hopping coefficient, and $V_j=V\cos(2\pi\alpha j+\theta)$ with $V$, $\theta$ and $\alpha$ being the quasiperiodic potential amplitude, the phase offset, and an irrational number, respectively. We note that the LSD pattern is independent of the specific values of $\alpha$ and $\theta$. This model undergoes the AT at $V=2J$, with all eigenstates being extended for $V < 2J$ and localized for $V > 2J$~\cite{AA}. For simplicity, we fix $J = 1$ and set $\alpha=F_{N-1}/F_{N}$ with $F_{N}$ being the Fibonacci sequence (i.e., $F_{1}=1$, $F_{2}=1$, and $F_{N}=F_{N-1}+F_{N-2}$). As $N$ approaches infinity, $\alpha$ converges to $(\sqrt{5}-1)/2$. Unless otherwise stated, we take the system size $L=F_{N}$, and use open boundary conditions.

We first compare the LSD of the localized phase in the AA model with that induced by random disorder, as shown in Fig.~\ref{01}. For the disorder-induced localization, we consider the above Eq.~(\ref{AAH}), with the onsite disorder $V_j$ being uniformly distributed in the interval $[-W, W]$. 
We observe that the energy spectrum of localized phases caused by disorder does not exhibit significant large gaps [Fig.~\ref{01}(a1)]. Apart from a decrease in the density of states (DOS) at the boundaries of the spectrum, the DOS across the spectrum is uniformly distributed [Fig.~\ref{01}(a2)]. As a contrast, the energy spectrum of quasiperiodic localized phases shows two distinct large gaps, dividing the spectrum into three segments [Fig.~\ref{01}(b1)]. The numbers of states in each segment from bottom to top are $F_{N-2}$, $F_{N-3}$, $F_{N-2}$, and at the boundaries of each segment, the DOS increases [Fig.~\ref{01}(b2)]. Then we compare the distribution of energy level spacings, defined as $\Delta E_n = E_{n+1}-E_n$, with the eigenvalues $E_n$ listed in ascending order. In the disorder system, the level statistics of localized phases are Poisson: $P(\delta E)=\frac{1}{\langle\delta E\rangle}exp(-\frac{\delta E}{\langle\delta E\rangle})$ [Fig.~\ref{01}(a3)], where $\delta E=\Delta E/\langle\Delta E\rangle$ and $\langle\delta E\rangle$ is the average of $\delta E$. Based on the energy level spacing, we can obtain the ratio of adjacent gaps as $r_n=\frac{min(\delta E_n, \delta E_{n+1})}{max(\delta E_n, \delta E_{n+1})}$~\cite{DAHuse1,DAHuse2}. For Poisson statistics, one can derive that the distribution of $r$ satisfies $P(r)=2/(1+r)^2$ [Fig.~\ref{01}(a4)], which gives the average value of $r$ as $\langle r\rangle=\int_0^1P(r)rdr=2\ln2-1\approx 0.387$. However, for the quasiperiodic localized phase, the energy level spacing noticeably deviates from Poisson statistics, as indicated by the black data points in Fig.~\ref{01}(b3).
The distribution $P(r)$ is not $2/(1+r)^2$ but rather takes on the form of a $\delta$ function [Fig.~\ref{01}(b4)].

\begin{figure}
	\centering
	\includegraphics[width=0.49\textwidth]{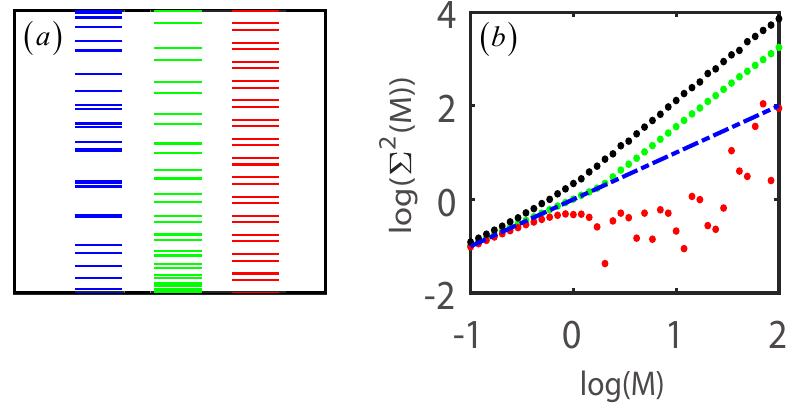}
	\caption{\label{02} (a) Examples of spectra. The blue uncorrelated energy levels are obtained in disordered AL phase.  
	The green and red energy levels respectively correspond to the edge and middle levels of the middle region in Fig.~\ref{01}(b1). Their corresponding averaged number variances are shown in (b). The green/red dots are calculated from the lowest/middle one-third spectrum of the middle region. The black dots and blue curve are calculated from the full spectrum in quasiperiodic and disordered systems. Here we take $30$ samples, with each sample specified by choosing an initial phase $\theta$. Other parameters are the same as those in Fig.~\ref{01}.}
\end{figure}

Before deriving the distributions $P(\delta E)$ and $P(r)$, we first investigate the uniformity of level spacings across different regions in the spectrum. Fig.~\ref{02}(a) displays three types of spectra, corresponding to the localized phase in disordered systems and the edge and middle regions in the middle segment of Fig.~\ref{01}(b1). The distances between the energy levels of the disordered system (blue lines) show significant fluctuations and lack correlation, allowing levels to approach each other arbitrarily closely. Similar properties are observed in the boundaries of the quasiperiodic system's energy spectrum (green lines). However, in the middle region of each segment of the energy spectrum, level repulsion is observed, expressing the unlikelihood of levels being degenerate in this system. This suggests that there is correlation in the energy spectrum of the localized phase in quasiperiodic systems. To characterize the uniformity of energy level spacings, we investigate the level number variance $\Sigma^2(\epsilon)$, defined as $\Sigma^2(\epsilon)=\langle M^2(\epsilon)\rangle-\langle M(\epsilon)\rangle^2$, where $\langle M(\epsilon)\rangle$ quantifies the average number of levels within the energy width $\epsilon$ on the unfolded scale~\cite{Dyson1963,Guhr1998,Bertrand2016,WangMBC}. In the unfolded spectrum, where the average spectral density is $1$, $\langle M(\epsilon) \rangle=\epsilon$~\cite{Bertrand2016,WangMBC}, thus $\epsilon$ can be replaced by $\langle M\rangle$, denoted by $M$ for simplicity. For Poisson statistics, the spectrum exhibits no correlations, resulting in a number variance that is exactly linear with a slope of one, i.e., $\Sigma^2(M)=M$ (blue dashed line in Fig.~\ref{02}(b)). 
Fig.~\ref{02}(a) shows that the distribution of energy levels in the middle region of each segment of the energy spectrum is more uniform, leading to a smaller $\Sigma^2$ (red dots in Fig.~\ref{02}(b)), similar to $\Sigma^2(M)\approx \frac{2}{\pi^2}\ln(2\pi M)$ that is obtained from the Wigner-Dyson distribution.  For each segment's boundary region (green dots in Fig.~\ref{02}(b)) and the overall energy spectrum (black dots in Fig.~\ref{02}(b)), when $M$ is large, meaning that the number of levels within the width $\epsilon$ is relatively high,
the linear slope of their respective $\Sigma^2$ is greater than $1$. This indicates that their energy level distribution is more uneven than the Poisson distribution. To the best of our knowledge, the spectra that satisfy the condition of the number variance being linear with a slope greater than $1$ have not been reported before. 

\begin{figure}
	\centering
	\includegraphics[width=0.49\textwidth]{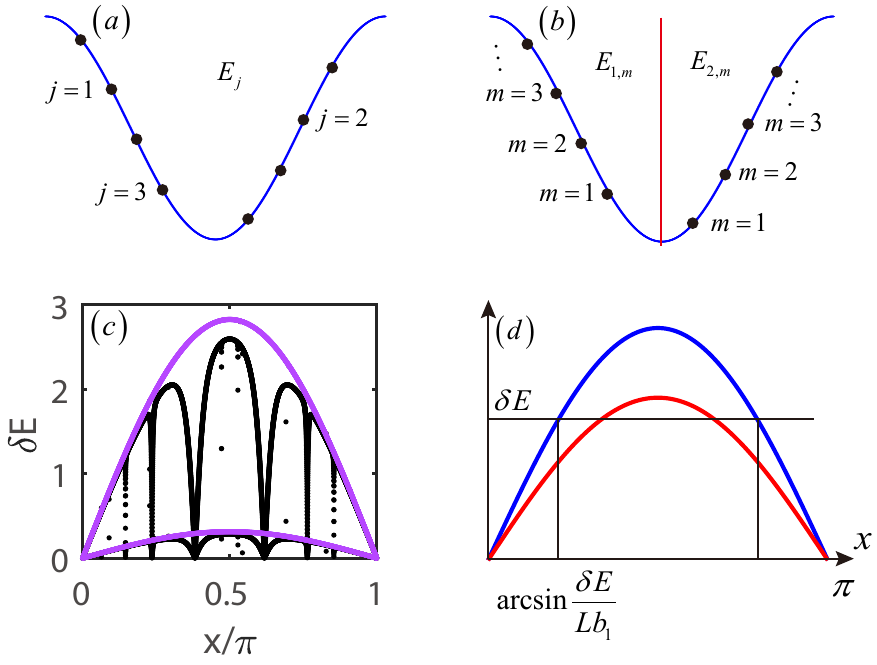}
	\caption{\label{03} Scheme of energy levels for (a) $E_{j}$ and (b) $E_{1,m}$ and $E_{2,m}$. (c) Level spacings of AA model with $V=10$, $\alpha=F_{N-1}/F_{N}$ and $L=28657$. The black and purple dots correspond to the hopping amplitudes $J=1$ and $J=0$, respectively.  (d) Scheme of level spacing as a function of $x$.}
\end{figure}

From the preceding discussion, one can see that the LSD of the AL phase induced by quasiperiodic potentials does not adhere to Poisson statistics. So, what type of statistical distribution does it exhibit?
We now deduce the LSD $P(\delta E)$ in the AA model's AL phase. We set $J=0$, $\alpha=F_{N-1}/F_N$, and fix $\theta$, then the system's eigenvalues are $E_{j}=V\cos(2\pi jF_{N-1}/F_{N}+\theta)$, with $j=0,1,2,\cdots,F_{N}-1$ [Fig.~\ref{03}(a)]. We introduce $n$, setting it equal to $jF_{N-1}$ mod($F_N$), then $E_{n}=-V\cos[2\pi n/F_{N}+\theta+\pi]$, and it is easy to verify that the range of $n$ is $n=0,1,2, ...,F_N-1$.
Shifting the labels of energies $n=m-n_{0}$ with $n_{0}=(\theta+\pi-\theta_{1})/(\frac{2\pi}{F_{N}})$, one can obtain
\begin{equation}
	E_{m}=-V\cos(2\pi\frac{m}{F_{N}}+\theta_{1}),~~~m=1,2,\cdots,F_{N}.
\end{equation}
By selecting the appropriate value for $n_0$, one can make the range of $\theta_1$ satisfy $\theta_{1}\in[-2\pi/F_{N},0)$~\cite{explain1}.
We then separate the energy levels into two parts, as shown in Fig.~\ref{03}(b). For $m=1,\cdots,F_{N}/2$, we denote the energies by $E_{1,m}$; for $m=F_{N}/2+1,F_{N}/2+2,\cdots,F_{N}$, we relabel $m\rightarrow F_{N}+1-m$ and denote the energies by $E_{2,m}$, hence $E_{1,m}=-V\cos(2\pi m/F_{N}+\theta_{1})$ and $E_{2,m}=-V\cos[2\pi (m-1)/F_{N}-\theta_{1}]$, with $m=1,2,\cdots,F_{N}/2$. It is convenient to introduce variables $x_{m}=2\pi(m-1/2)/F_{N}$ and $y=\pi/F_{N}+\theta_{1}\in[-\pi/F_{N},\pi/F_{N})$, and then the energies become
\begin{equation}\label{E1E2}
	\begin{cases}
		E_{1}(x_{m})=-V\cos(x_m+y), \\
		E_{2}(x_{m})=-V\cos(x_m-y),
	\end{cases}
\end{equation}

The energies are naturally ordered:
\begin{equation}
	E_{1}(x_{m})<E_{1}(x_{m+1}),~~E_{2}(x_{m})<E_{2}(x_{m+1}).
\end{equation}
If $0< y\leq\frac{\pi}{F_{N}}$, the total energies are ordered by $E_{2}(x_{1})<E_{1}(x_{1})<E_{2}(x_{2})<E_{1}(x_{2})<\cdots$, thus $\Delta E_{1}(x_{m})=E_{1}(x_{m})-E_{2}(x_{m})$ and $\Delta E_{2}(x_{m})=E_{2}(x_{m+1})-E_{1}(x_{m})$. 
Combining Eq.~(\ref{E1E2}), we can obtain that
\begin{equation}\label{xy}
		\begin{cases}
			\Delta E_{1}(x_{m})=2Vb_{1}\sin(x_{m}), \\
			\Delta E_{2}(x_{m})=2Vb_{2}\sin(x_{m}+\frac{\pi}{F_{N}}).
		\end{cases} \\
\end{equation}
where $b_{1}=\sin y$ and $b_{2}=\sin(\pi/F_{N}-y)$. When $-\pi/F_{N}<y<0$, $E_{1}(x_{1})<E_{2}(x_{1})<E_{1}(x_{2})<E_{2}(x_{2})<\cdots$, one can obtain Eq.~(\ref{xy}), and it still holds true, with the only difference being that $b_{1}=-\sin y$ and $b_2=\sin(\pi/F_{N}+y)$. 

For the limit $L\rightarrow\infty$, we set $x_{m}\rightarrow x$. We then consider $\delta E_{1}(x_{m})=\frac{\Delta E_{1}(x_{m})}{\langle\Delta E_{m}\rangle}=\frac{\Delta E_{1}(x_{m})}{2V/L}$, and combining Eq.~(\ref{xy}), we obtain
\begin{equation}\label{b1}
\delta E_{1}(x_{m})=b_1L\sin(x_{m}).
\end{equation}
Similarly, 
\begin{equation}\label{b2}
	\delta E_{2}(x_{m})=b_{2}L\sin(x_{m}+\frac{\pi}{F_{N}}).
\end{equation}
We note that $b_1$ and $b_2$ are of the order of $1/F_N$, so $b_1L$ and $b_2L$ are of the order of $1$.
Therefore, the distribution of $\delta E$ consists of two branches, as shown in Fig.~\ref{03}(c), and these two branches satisfy Eq.~(\ref{b1}) and Eq.~(\ref{b2}), respectively.
Then one can calculate that the total number of states for the energy smaller than $\delta E$ is $N_{P}(\delta E_{1}\leq\delta E)=2N_P(0\leq x\leq\arcsin(\frac{\delta E}{b_{1}L}))=\frac{2\arcsin(\frac{\delta E}{b_{1}L})}{\pi}$ [see Fig.~\ref{03}(d)].
Hence the probability distribution is
\begin{equation}\label{limt}
	P_{1}(\delta E)=\frac{dN_{P}(\delta E_{1}\leq\delta E)}{d\delta E}=\frac{2}{\pi\sqrt{(b_{1}L)^2-\delta E^2}}.
\end{equation}
Similarly, one can obtain $P_{2}(\delta E)=\frac{2}{\pi\sqrt{(b_{2}L)^2-\delta E^2}}$. The total probability distribution should be considered as the sum of the two. 

\begin{figure}
	\centering
	\includegraphics[width=0.49\textwidth]{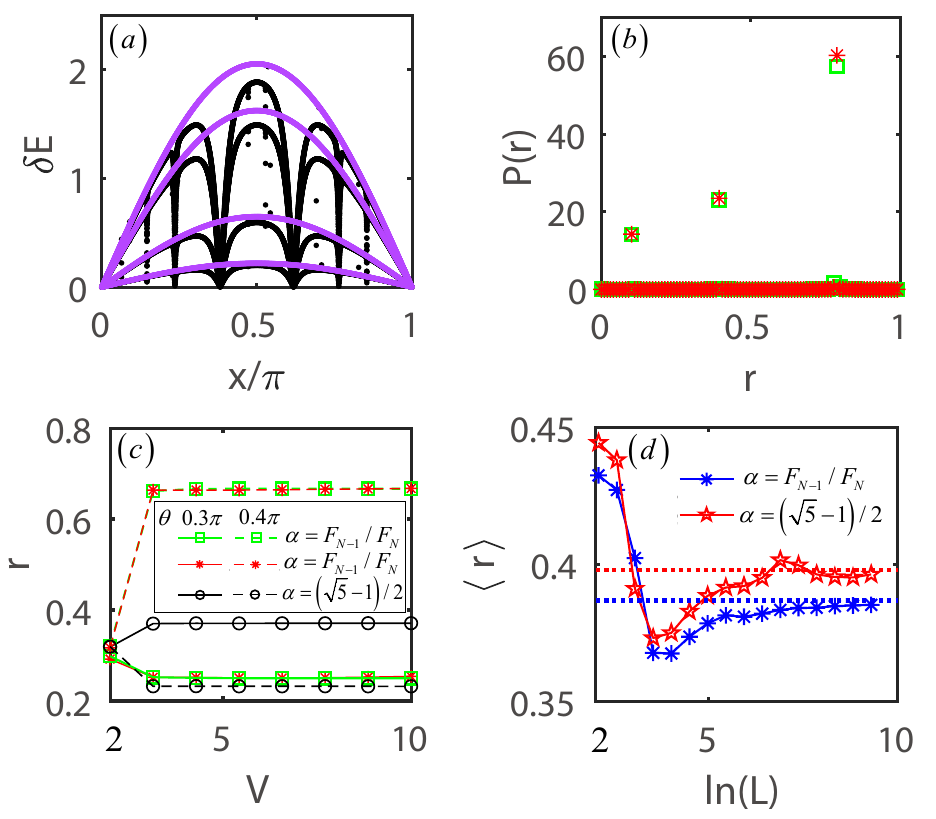}
	\caption{\label{04} (a) level spacing distributions and (b) $P(r)$ of AA model with $\alpha=(\sqrt{5}-1)/2$. Other parameters are the same as those in Fig.~\ref{03}(c) and Fig.~\ref{01}(b4). (c) For $\alpha=(\sqrt{5}-1)/2$ and $\alpha=F_{N-1}/F_{N}$, $r$ varies with different initial phases $\theta$ and the strength of the quasiperiodic potential $V$. The green/red dots are calculated from the lowest/middle one-third of the spectrum in the central region [see Fig.~\ref{01}(b1)], and the black dots are obtained from the entire energy spectrum. (d) For $\alpha=(\sqrt{5}-1)/2$ and $\alpha=F_{N-1}/F_{N}$, $\langle r\rangle$ averaged over $30$ samples and all energy levels varies with changes in size.}
\end{figure}

We previously discussed the case of $V\gg J$. For the general case, it is challenging to derive its expression. 
From Eq.~(\ref{limt}), it can be seen that when $\delta E=b_1L$, the distribution of $\delta E$ is divergent. From Eq.~(\ref{b1}), it is evident that the point where $P(\delta E)$ diverges is the maximum of each branch of the $\delta E$ distribution. When $J=0$, there are two branches of $\delta E$. Therefore, as $\delta E$ increases, $P(\delta E)$ will diverge twice [purple dashed lines in Fig.~\ref{01}(b3)]. However, in general, there are more than two branches of $\delta E$ [black dots in Fig.~\ref{03}(c)]. Hence,
we speculate that for AL induced by quasiperiodic potential, the LSD in the general case should satisfy a unified form 
\begin{equation}\label{pdeltaE}
	P(\delta E)=\sum_{\kappa}[\frac{A_{\kappa}}{\pi\sqrt{(b_{\kappa}L)^2-\delta E^2}}+C_{\kappa}]\Theta((b_{\kappa}L)^2-\delta E^2),
\end{equation}
where $\Theta$ is the step function~\cite{explaintheta}, $\kappa\geq2$ represents the number of branches, $A_{\kappa}$ and $C_{\kappa}$ are the undetermined parameters that describe the scaling and translation of $P(\delta E)$ with $J=0$.
Here we introduce parameters ($A_{\kappa}, b_{\kappa}, C_{\kappa}$) that depend on the strength of the quasiperiodic potential, referencing the statistically unified form $P(\delta E)=Aexp(-B\frac{\delta E}{\langle\delta E\rangle})$ for the LSD induced by disorder, with the fitting parameters $A$ and $B$ changing with increasing disorder strength. We note that although there is a summation over $\kappa$ in Eq.~(\ref{pdeltaE}), the divergence behavior of $P(\delta E)$ is determined by the vicinity of the maximum value of each branch of $\delta E$. The influence of other branches is minimal. Therefore, without summation, Eq.~(\ref{pdeltaE}) can still fit the distribution of $P(\delta E)$ well [light blue dashed lines in Fig.~\ref{01}(b3)].


We further derive the distribution of the adjacent gap ratio $r$ through the use of the defining equation $P(r) = \int d(\delta E_{n},\delta E_{n+1})\delta(r-\frac{\min\{\delta E_{n},\delta E_{n+1}\}}{\max\{\delta E_{n},\delta E_{n+1}\}})p(\delta E_{n},\delta E_{n+1})$. For the case of $J=0$, $\delta E_{n}$ and $\delta E_{n+1}$ respectively correspond to the two purple lines in Fig.~\ref{03}(c), which are described by Eq.~(\ref{b1}) and Eq.~(\ref{b2}). Considering $\pi/F_N\rightarrow 0$, 
$\delta(r-\frac{\min\{\delta E_{n},\delta E_{n+1}\}}{\max\{\delta E_{n},\delta E_{n+1}\}})=\delta(r-\frac{\min\{b_{1},b_{2}\}}{\max\{b_{1},b_{2}\}})$ can be brought outside the integral, so
\begin{equation}\label{Pr}
		P(r) = \delta(r-\frac{\min\{b_{1},b_{2}\}}{\max\{b_{1},b_{2}\}}).
\end{equation}
Thus, the distribution $P(r)$ is $\delta$ function, as shown in Fig.~\ref{01}(b4), which is clearly different from the $P(r)$ given by Poisson statistics. When $\alpha=(\sqrt{5}-1)/2$, the increasing order of crossings between $E_{1,m}$ and $E_{2,m}$ as $m$ increases shown in Fig.~\ref{03}(b) will be disrupted (see Appendix), which leads to
the number of branches of $\delta E$ exceeding $2$, as shown in Fig.~\ref{04}(a). Consequently, multiple peaks appear in $P(r)$, as depicted in Fig.~\ref{04}(b). 
For the case of $J\neq 0$, using perturbation theory, we can demonstrate that $b_1$ and $b_2$ in Eq.~(\ref{xy}) need to be multiplied by the same factor~(see Appendix). Therefore, from Eq.~(\ref{Pr}), $P(r)$ is independent of both $J$ and $V$ in the AL phase. Additionally, the expressions for $b_1$ and $b_2$ include the initial phase $\theta$, hence the peak positions of $P(r)$ depend on $\theta$. From Fig.~\ref{04}(c), we observe that the values of $r$ are independent of $V$ and the position in the energy spectrum. However, they depend on $\theta$ and on whether $\alpha$ takes the value $(\sqrt{5}-1)/2$ or $F_{N-1}/F_N$. 

Then we consider the sample average of $r$, which is equivalent to average over $y$. We suppose $0< y\leq\frac{\pi}{F_{N}}$, then $\langle r\rangle = \frac{F_{N}}{\pi}\int_{0}^{\pi/F_N} dy \frac{\min\{b_{1},b_{2}\}} {\max\{b_{1},b_{2}\}}$. When $L=F_N\rightarrow\infty$, $b_{1}=\sin y\sim y$ and $b_{2}=\sin(\pi/F_{N}-y)\sim \pi/F_{N}-y$, so 
$\langle r\rangle= \frac{F_{N}}{\pi}(\int_{0}^{\frac{\pi}{2F_{N}}}dy\frac{y}{\frac{\pi}{F_{N}}-y} +\int_{\frac{\pi}{2F_{N}}}^{\frac{\pi}{F_{N}}}dy\frac{\frac{\pi}{F_{N}}-y}{y}) = 2\ln2-1$, as shown in Fig.~\ref{04}(d). When  $\frac{-\pi}{F_{N}}<y\leq0$, one can easily obtain the same result. We note that although the result of $\langle r\rangle$ is the same as that given by Poisson distribution, it is not caused by Poisson statistics. When $\alpha=(\sqrt{5}-1)/2$, we mentioned earlier that
the distribution of $P(r)$ has multiple peaks, which is different from the case of $\alpha=\frac{F_{N-1}}{F_{N}}$, where there is only a single peak. Naturally, the aforementioned process of calculating $\langle r\rangle$ is no longer applicable for the case of $\alpha=(\sqrt{5}-1)/2$.  Our numerical results
show that it is close to $0.4$, distinct from $0.387$ [Fig.~\ref{04}(d)]. 

When the interaction is added, even with a fixed $\theta$, the LSD of the quasiperiodic MBL phase still follows Poisson statistics~(see Appendix), 
which is consistent with the results of previous studies~\cite{Khemani2017,Roushan2017,XiaopengLi,Modak2015,Vu2022}. In general, the LSD of both the AL and MBL phases induced by disorder obey Poisson statistics. The LSD of the MBL phase induced by quasiperiodic potentials also conforms to Poisson statistics. However, the LSD of the AL induced by quasiperiodic potentials does not follow Poisson statistics.

\section{Conclusion and Discussion}
We have derived the LSD of the quasiperiodic AL phase, satisfying Eq.~(\ref{pdeltaE}), which does not follow Poisson statistics. In addition, we found more differences in the spectrum between the quasiperiodic and disordered AL phases. Specifically: (1) the former exhibits different degrees of uniformity in level spacing across different spectral regions, and the overall distribution is even more uneven than a Poisson distribution, whereas the latter shows a relatively uniform level distribution across different spectral regions; (2) The distribution of $P(r)$ for the former is a $\delta$ function dependent only on the initial phase, while the distribution of $P(r)$ for the latter follows $P(r)=2/(1+r)^2$; (3) The sample-averaged value $\langle r\rangle$ for the former depends on whether $\alpha=F_{N-1}/F_N$ or $\alpha=(\sqrt{5}-1)/2$. Although, for the case of $\alpha=F_{N-1}/F_N$, the obtained $\langle r\rangle$ is the same as that obtained from Poisson statistics, it does not originate from Poisson statistics. Further, there are spatial correlations in quasiperiodic systems, indicating that increasing the number of samples is not equivalent to increasing the size, which is in contrast to Poisson statistics. Thus, for the quasi-periodic Anderson localized phase, there is no physical basis for sampling averaging over $r$. The energy spectrum of quasiperiodic systems can be experimentally determined in various systems, such as semiconductor quantum dots~\cite{Kiczynski2022} or superconducting qubits~\cite{Roushan2017}.

\begin{acknowledgments}
We thank Qi Zhou for making us aware of Ref.~\cite{Jitomirskaya}. This work is supported by National Key R\&D Program of China under Grant No.2022YFA1405800, the National Natural Science
Foundation of China (Grant No.12104205), the Key-Area Research and Development Program of Guangdong Province (Grant No. 2018B030326001), Guangdong Provincial Key Laboratory (Grant No.2019B121203002).
\end{acknowledgments}

\appendix


\section{Level-spacing distribution of AA model with $\alpha=(\sqrt{5}-1)/2$}
In this section we show the LSD of the AA model for fixed $\theta$ in the limit $J=0$. The energies are
\begin{equation}\label{1.1}
	E_{j}^{(0)}=V\cos(2\pi\alpha j+\theta),~~~~~~~~j=0,1,2,\cdots,F_{N}-1, 
\end{equation}
with $\alpha=\frac{\sqrt{5}-1}{2}$ and $\theta\in[-\pi,\pi)$. We separate $\alpha$ into two parts:
\begin{equation}\label{1.2}
	\alpha=\frac{F_{N-1}}{F_{N}}+\beta_{N}\equiv\alpha_{N}+\beta_{N},
\end{equation}
with $\alpha_{N}=\frac{F_{N-1}}{F_{N}}$ is the ratio of Fibonacci sequence. As the Fibonacci numbers $F_N$ can be expressed by $F_{N}=\frac{1}{\sqrt{5}}[\alpha^{-N}-(-\alpha)^{N}]$, hence in the large $N$ limit, $\beta_{N}$ can be evaluated by
\begin{eqnarray}\label{1.3}
	\beta_{N}=\alpha-\frac{F_{N-1}}{F_{N}}= \alpha-\frac{\alpha^{1-N}-(-\alpha)^{N-1}}{\alpha^{-N}-(-\alpha)^{N}} \nonumber\\
	\sim  \frac{1+\alpha^2}{\alpha}(-1)^{N+1}\alpha^{2N},\qquad \qquad \qquad
\end{eqnarray}
so $\beta_{N}\rightarrow 0$ exponentially as $N\rightarrow\infty$. Hence the energies in Eq. \ref{1.1} can be represented as
\begin{equation}\label{1.4}
	E_{j}^{(0)} = -V\cos(2\pi\frac{\gamma_j}{F_{N}}+2\pi\beta_{N}j+\theta+\pi).
\end{equation}
with $\gamma_j=jF_{N-1}(\mod~F_{N})$. Define $n=jF_{N-1}(\mod~F_{N})$, then we have $j=nF_{N-1}(\mod~F_{N})$ for even $N$. Hence the energies can be relabeled by
\begin{eqnarray}\label{1.5}
	E_{n}^{(0)}=-V\cos[2\pi\frac{n}{F_{N}}+2\pi\beta_{N}\gamma_{n}+\theta+\pi],\nonumber\\
	~~~~~n=0,1,\cdots,F_{N}-1.
\end{eqnarray}

Shifting the labels of energies $n=m-n_{0}$ with
\begin{equation}\label{1.6}
	n_{0}=\frac{\theta+\pi-\theta_{1}}{\frac{2\pi}{F_{N}}},~~~~~~~~n_{0}=1,\cdots,F_{N},
\end{equation}
where $\theta_1$ is
\begin{equation}\label{1.7}
	\theta_{1}=[\theta+\pi](\mod~\frac{2\pi}{F_{N}})-\frac{2\pi}{F_{N}},~~~~~\theta_{1}\in[-\frac{2\pi}{F_{N}},0).
\end{equation}
Then we see
\begin{eqnarray}\label{1.8}
	E_{m}^{(0)}=-V\cos(2\pi\frac{m}{F_{N}}+2\pi\beta_{N}\gamma_{m-n_{0}}+\theta_{1}),\nonumber\\
	~~~m=1,2,\cdots,F_{N}.
\end{eqnarray}
Now separating the energies levels into two parts. For $m=1,\cdots,\frac{F_{N}}{2}$ we denote the energies by $E_{1,m}^{(0)}$; for $m=\frac{F_{N}}{2}+1,\frac{F_{N}}{2}+1,\cdots,F_{N}$ we relabel $m\rightarrow F_{N}+1-m$ and denote the energies by $E_{2,m}^{(0)}$, hence
\begin{equation}\label{1.9}
	\begin{cases}
		E_{1,m}^{(0)}=-V\cos(2\pi\frac{m}{F_{N}}+2\pi\beta_{N}\gamma_{m-n_{0}}+\theta_{1}), \\
		E_{2,m}^{(0)}=-V\cos(2\pi\frac{m-1}{F_{N}}-2\pi\beta_{N}\gamma_{1-m-n_{0}}-\theta_{1}),
	\end{cases}
\end{equation}
with $m=1,2,\cdots,\frac{F_{N}}{2}$. It is convenient to define the variable $x_{m}=2\pi\frac{m-\frac{1}{2}}{F_{N}}$, $y=\frac{\pi}{F_{N}}+\theta_{1}\in[-\frac{\pi}{F_{N}},\frac{\pi}{F_{N}})$ and $z_{m}=2\pi\beta_{N}\gamma_{m}$, the energies are
\begin{equation}\label{1.10}
	\begin{cases}
		E_{1}^{(0)}(x_{m})=-V\cos(x_m+y+z_{m-n_{0}}), \\
		E_{2}^{(0)}(x_{m})=-V\cos(x_m-y-z_{1-m-n_{0}}).
	\end{cases}
\end{equation}

Now we first consider the limit $\beta_N=0$ $(z=0)$ and suppose $F_{N}$ is even for simplification. The energies are naturally ordered:
\begin{equation}\label{1.11}
	E_{1}^{(0)}(x_{m})<E_{1}^{(0)}(x_{m+1}),~~~~E_{2}^{(0)}(x_{m})<E_{2}^{(0)}(x_{m+1}).
\end{equation}
Hence the full energies is ordered by
\begin{widetext}
	\begin{equation}\label{1.12}
		\begin{cases}
			if &~~0\leq y\leq\frac{\pi}{F_{N}},~~~~~~~E_{2}^{(0)}(x_{1})<E_{1}^{(0)}(x_{1})<E_{2}^{(0)}(x_{2})<E_{1}^{(0)}(x_{2})<\cdots. \\ if &~~ -\frac{\pi}{F_{N}}<y<0,~~~~E_{1}^{(0)}(x_{1})<E_{2}^{(0)}(x_{1})<E_{1}^{(0)}(x_{2})<E_{2}^{(0)}(x_{2})<\cdots.
		\end{cases}
	\end{equation}
\end{widetext}
The level spacing for $0\leq y\leq\frac{\pi}{F_{N}}$ satisfy
\begin{widetext}
	\begin{equation}\label{1.13}
		\begin{split}
			if & ~~~~0\leq y\leq\frac{\pi}{F_{N}},~~~~~~~~
			\begin{cases}
				\Delta E_{1}^{(0)}(x_{m})=E_{1}^{(0)}(x_{m})-E_{2}^{(0)}(x_{m}), \\
				\Delta E_{2}^{(0)}(x_{m})=E_{2}^{(0)}(x_{m+1})-E_{1}^{(0)}(x_{m}).
			\end{cases} \\
			if & ~~~~-\frac{\pi}{F_{N}}<y<0,~~~~
			\begin{cases}
				\Delta E_{1}^{(0)}(x_{m})=E_{2}^{(0)}(x_{m})-E_{1}^{(0)}(x_{m}), \\
				\Delta E_{2}^{(0)}(x_{m})=E_{1}^{(0)}(x_{m+1})-E_{2}^{(0)}(x_{m}).
			\end{cases}
		\end{split}
	\end{equation}
\end{widetext}
After a straightforward calculation, we can get
\begin{equation}\label{1.14}
	\begin{cases}
		\delta E_{1}^{(0)}(x_{m})=b_{1}F_{N}\sin(x_{m}), \\
		\delta E_{2}^{(0)}(x_{m})=b_{2}F_{N}\sin(x_{m}+\frac{\pi}{F_{N}}).
	\end{cases}
\end{equation}
with $\delta E=F_{N}\Delta E/2V$. We see that the level spacings lie on two sine functions with amplitude $2Vb_{1/2}$ and
\begin{equation}\label{1.15}
	\begin{split}
		if & ~0\leq y\leq\frac{\pi}{F_{N}},~~b_{1}=\sin y,~~b_{2}=\sin(\frac{\pi}{F_{N}}-y), \\
		if & ~-\frac{\pi}{F_{N}}<y<0,~~b_{1}=-\sin y,~~b_{2}=\sin(\frac{\pi}{F_{N}}+y).
	\end{split}
\end{equation}
We see the amplitudes satisfy
\begin{equation}\label{1.16}
	b_{1}+b_{2}=\frac{\pi}{F_{N}}+O(F_{N}^{-3}).
\end{equation}

\begin{figure}[htbp]
	\centering
	\includegraphics[width=1\linewidth]{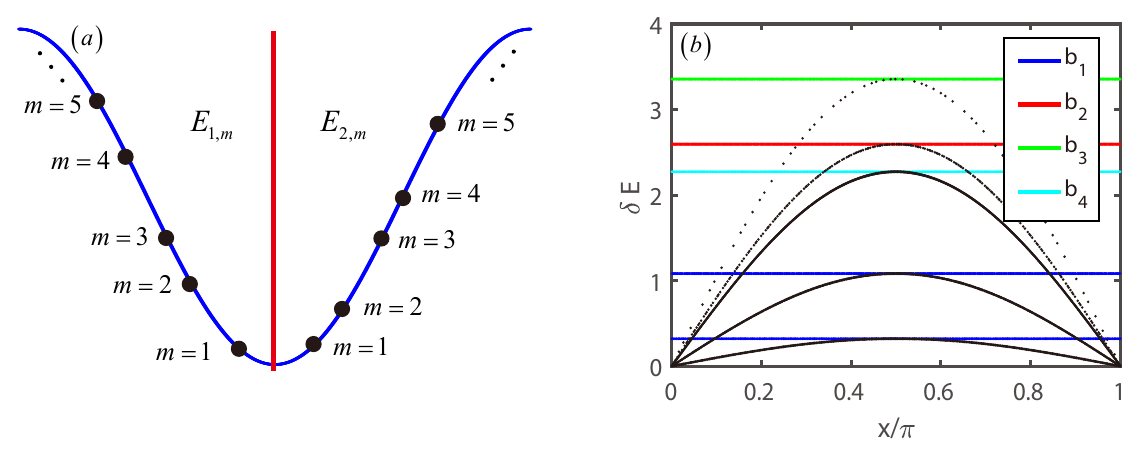}\\
	\caption{(a)Scheme of energies levels for non vanishing $\beta_{N}$. (b)Level spacing for $\alpha=\frac{\sqrt{5}-1}{2}$. The amplitude $b_j$s are calculated from Eq. \ref{1.18}.}\label{FigS1}
\end{figure}

For the case $\beta_{N}\ne0$, there is another phase shift $z_m=2\pi\beta_{N}\gamma_m$ in Eq. \ref{1.9} except $\theta_1$. In the large $N$ limit, the Fibonacci number $F_{N}=\frac{\alpha^{-N}}{\sqrt{5}}$, then the phase shift can be evaluated from Eq. \ref{1.3} that $|2\pi\beta_{N}F_{N}|\sim\frac{1+\alpha^2}{5\alpha}(-1)^{N+1}\frac{2\pi}{F_{N}}\sim0.8944\frac{\pi}{F_{N}}$, which is in the same order of $\frac{\pi}{F_{N}}$. Hence the order of energies Eq. \ref{1.11} remains while the order of full energies Eq. \ref{1.12} can break as shown in Fig. \ref{FigS1}(a). Here we show an example for $\theta=0.4\pi$ and $N=18$. Then we can derive $n_0=1809$ and $y=0.6\frac{\pi}{F_N}$. All the energy level spacings lie on
\begin{widetext}
	\begin{equation}\label{1.17}
		\begin{split}
			|\Delta E_1| =& |E_{1}(x_m)-E_{2}(x_m)|=2Vb_1\sin(x_m+\frac{z_{m-n_0}+z_{1-m-n_0}}{2}), \\
			|\Delta E_2| =& |E_{2}(x_{m+1})-E_{1}(x_m)|=2Vb_2\sin(x_m+\frac{\pi}{F_N}+\frac{z_{m-n_0}+z_{-m-n_0}}{2}), \\ |\Delta E_3| =& |E_{1}(x_{m+1})-E_{2}(x_m)|=2Vb_3\sin(x_m+\frac{\pi}{F_N}+\frac{z_{1+m-n_0}-z_{1-m-n_0}}{2}), \\ |\Delta E_4| =& |E_{1}(x_{m+1})-E_{1}(x_m)|=2Vb_4\sin(x_m+y+\frac{z_{1+m-n_0}+z_{m-n_0}}{2}), \\ |\Delta E_5| =& |E_{2}(x_{m+1})-E_{2}(x_m)|=2Vb_5\sin(x_m+\frac{\pi}{F_N}-y-\frac{z_{-m-n_0}+z_{1-m-n_0}}{2}),
		\end{split}
	\end{equation}
\end{widetext}
where the amplitude
\begin{equation}\label{1.18}
	\begin{split}
		b_1 =& |\sin(y+\frac{z_{m-n_0}+z_{1-m-n_0}}{2})|, \\
		b_2 =& |\sin(\frac{\pi}{F_{N}}-y-\frac{z_{-m-n_0}+z_{m-n_0}}{2})|, \\
		b_3 =& |\sin(\frac{\pi}{F_{N}}+y+\frac{z_{1+m-n_0}+z_{1-m-n_0}}{2})|, \\
		b_4 =& |\sin(\frac{\pi}{F_{N}}+\frac{z_{1+m-n_0}-z_{m-n_0}}{2})|, \\
		b_5 =& |\sin(\frac{\pi}{F_{N}}+\frac{z_{1-m-n_0}-z_{-m-n_0}}{2})|.
	\end{split}
\end{equation}
It can be seen that $b_{4}=b_{5}$ is independent of $y$ and $\theta$. From the identity
\begin{equation}\label{1.19}
	z_{m+a}+z_{-m+b}=2\pi\beta_{N}kF_{N}+z_{a+b},~~k=1,2,
\end{equation}
where $z_{a+b}$ is a constant independent of $m$, so each $b_j$ has two values, and the amplitude can be several of these $b_j$s. In Fig. \ref{FigS1}(b) we show the level spacing as a function of $x$, and see that there are five relative $b_j$s. As the adjacent gap $P(r)$ is the ratio of nearest level spacing, more delta peaks emerge as shown in the main text. After averaging over $y$, the mean adjacent gap is around $0.4$ from numerical simulation.

\section{Energies levels of AA model from perturbative perspective}
\begin{figure*}[htbp]
	\centering
	\includegraphics[width=0.6\linewidth]{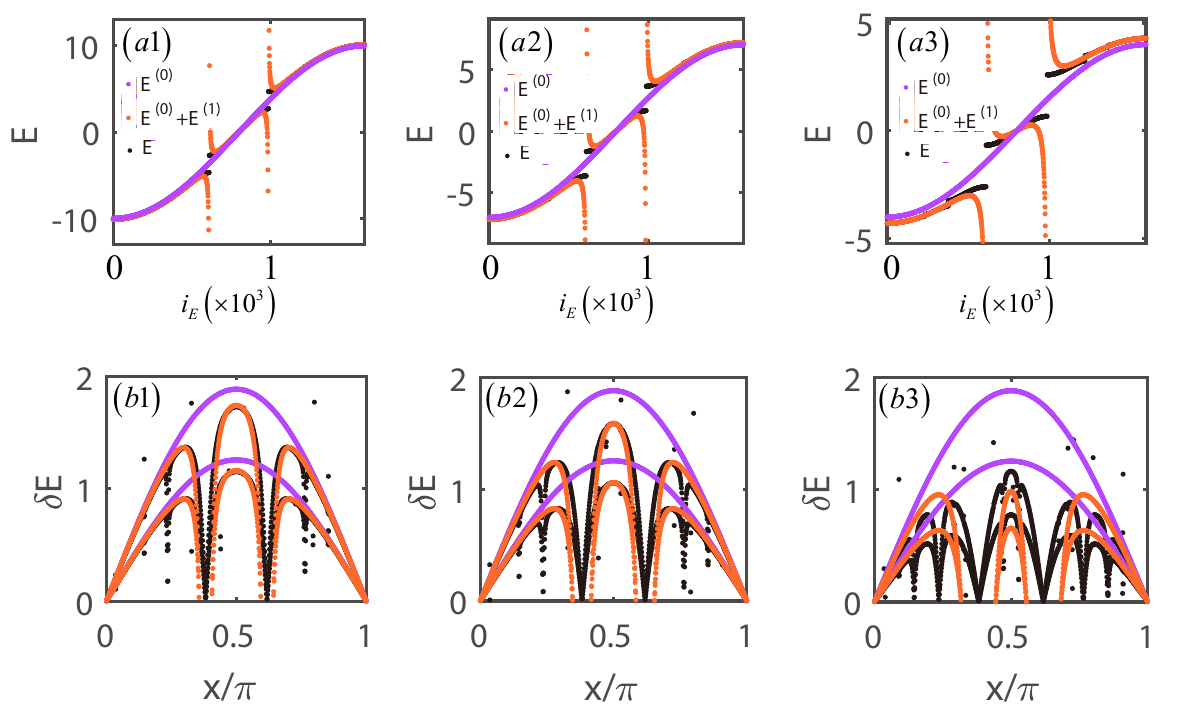}\\
	\caption{(a1)-(a3)The perturbative energy levels. (b1)-(b3)The perturbative level spacings. The hopping amplitude $J=1$, the phase factor $\theta=0.2\pi$, and the system size $L=1597$. The strength of qausi-periodic potential is $V=10$ for (a1)(b1), $V=7$ for (a2)(b2), and $V=4$ for (a3)(b3) respectively.}\label{FigS2}
\end{figure*}

In this section, we discuss the energies levels for finite hopping amplitude from perturbation theory. Separating the Hamiltonian into two parts
\begin{equation}\label{2.1}
	H=H_{J}+H_{V},
\end{equation}
with $H_{J/V}$ denoting the kinetic/potential energies with $|H_{J}|\ll|H_{V}|$. Up to the second order of $J/V$, the energies are
\begin{equation}\label{2.2}
	E_{j} = E_{j}^{(0)}+\sum_{l\ne j}\frac{|\langle j|H_{J}|l\rangle|^2}{E_{j}^{(0)}-E_{l}^{(0)}} = E_{j}^{(0)}+\frac{J^2E_{j}^{(0)}}{[E_{j}^{(0)}]^2-V^2\cos^2(\pi\alpha)}.
\end{equation}
where $E_{j}^{(0)}$ is the unperturbated energy in Eq. \ref{1.1}. We see two gaps are opened at $E_{gap}=\pm V\cos(\pi\alpha)$ and the perturbation breaks down near the band edge. Near the center the band $E_{j}\sim0$, the energies are renormalized by $1-\frac{J^2}{V^2\cos^2(\pi\alpha)}$. In Fig. \ref{FigS2}(a1)-(a3) we show energies levels for different $V$. The perturbation works well near the center of the band even for small qausi-periodic potential.

Now we move to the behavior of level spacing. Substituting Eq. \ref{2.2} into Eq. \ref{1.13} we see the leading order of level spacing are
\begin{equation}\label{2.3}
	\begin{split}
		\delta E_{1}(x_m) =& b_{1}F_{N}\sin x_{m}\bigg[1-\frac{J^2}{V^2}\frac{\cos^2(\pi\alpha)+\cos^2 x_m}{[\cos^2(\pi\alpha)-\cos^2 x_m]^2}\bigg], \\ \delta E_{2}(x_m) =& b_{2}F_{N}\sin x_{m}\bigg[1-\frac{J^2}{V^2}\frac{\cos^2(\pi\alpha)+\cos^2 x_m}{[\cos^2(\pi\alpha)-\cos^2 x_m]^2}\bigg].
	\end{split}
\end{equation}
We see that the level spacing is modified by a function dependent on $x_m$, and could divergence near $\cos x_{m}=\pm\cos(\pi\alpha)$, but the ratio $\frac{\delta E_{1}}{\delta E_{2}}=\frac{b_1}{b_2}$ is a constant for fixed phase. Hence the adjacent gap $P(r)$ is almost $J$ independent. In Fig. \ref{FigS2}(b1)-(b3) we show the level spacing as a function of $x$, we see that the perturbation works well even near the band edge.

\begin{figure}[h]
	\centering
	\includegraphics[width=0.9\linewidth]{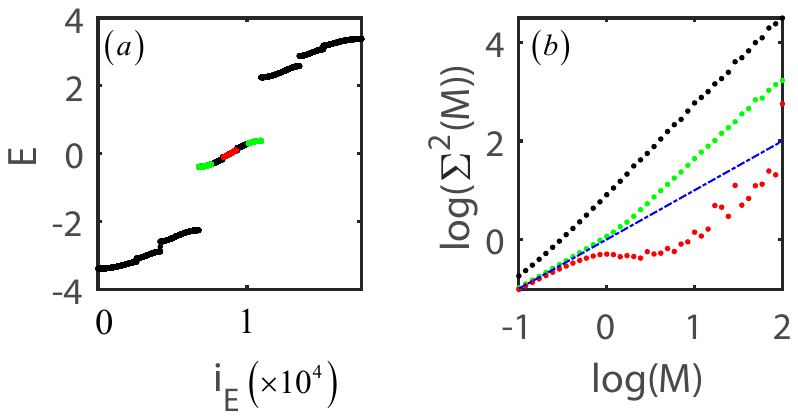}\\
	\caption{(a)Energy level of AA model for system size $L=F_{22}=17711$. The parameters $V/J=3$ and $\theta=0.3\pi$. The green/red/green curves correspond to the lowest/middle/highest $1/5$ energies of the middle region. (b)The green/red dots are calculated from the lowest/middle $1/5$ spectrum of the middle region. The black dots and blue curves are calculated from the full spectrum in quasiperiodic and disordered systems.}\label{S03}
\end{figure}

In Fig. 2 of the main text, we have shown the number variance $\Sigma^2(M)$ for $V/J=10$, which exhibits quantitatively different behaviors from uncorrelated energy levels. The Fig. \ref{S03} shows the case of $V/J=3$. We see that it does not have qualitative differences compared to $V/J=10$.

\begin{figure}[h]
	\centering
	\includegraphics[width=0.3\textwidth]{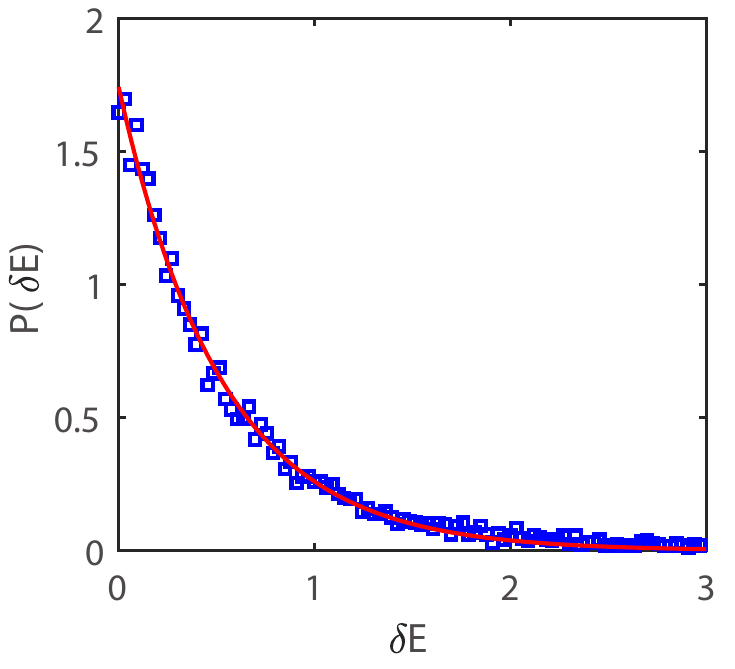}
	\caption{\label{05} The distribution of adjacent gap for $J=1$, $V=5$, $\theta=0.3\pi$ and $U=2$. The system size $L=16$ and the number of particles $N=8$. The fitting function $P(\delta E)=1.742e^{-1.903\delta E}$.}
\end{figure}

\section {Level statistics of many-body quasiperiodic systems}
On the AA model, we introduce nearest-neighbor interactions $U\sum_jn_jn_{j+1}$ with a fixed initial phase $\theta$. From Fig. \ref{05}, it can be observed that its level-spacing distribution follows Poisson statistics. Therefore, the localized phases in many-body quasiperiodic systems exhibit Poisson statistics.



\begin{thebibliography}{36}
\bibitem{Anderson1958} P. W. Anderson, Phys. Rev. {\bf 109}, 1492 (1958). 
\bibitem{RMP1} P. A. Lee and T. V. Ramakrishnan, Disordered electronic systems, Rev. Mod. Phys. {\bf 57}, 287 (1985).
\bibitem{RMP2} F. Evers and A. D. Mirlin, Anderson transitions, Rev. Mod. Phys. {\bf 80}, 1355 (2008).
\bibitem{Kramer1993} B. Kramer and A. MacKinnon, Localization: theory and experiment, Rep. Prog. Phys. {\bf 56}, 1469 (1993).

\bibitem{Soukoulis1981} C. M. Soukoulis and E. N. Economou, Localization in One-Dimensional Lattices in the Presence of Incommensurate Potentials, Phys. Rev. Lett. {\bf 48}, 1043 (1981). 
\bibitem{DasSarma1988} S. Das Sarma, S. He, and X. C. Xie, Mobility edge in a model one-dimensional potential, Phys. Rev. Lett. {\bf 61}, 2144 (1988).
\bibitem{Biddle2009} J. Biddle, B. Wang, D. J. Priour Jr, and S. Das Sarma, Localization in one-dimensional incommensurate lattices beyond the Aubry-André model, Phys. Rev. A {\bf 80}, 021603 (2009); J. Biddle and S. Das Sarma, Predicted mobility edges in one-dimensional incommensurate optical lattices: an exactly solvable model of Anderson localization, Phys. Rev. Lett. {\bf 104}, 070601 (2010).
\bibitem{XLi2017} X. Li, X. Li, and S. Das Sarma, Mobility edges in one dimensional bichromatic incommensurate potentials, Phys. Rev. B {\bf 96}, 085119 (2017);  D. Vu and S. Das Sarma, Generic mobility edges in several classes of duality-breaking one-dimensional quasiperiodic potentials, Phys. Rev. B {\bf 107}, 224206 (2023).
\bibitem{HYao2019} H. Yao, H. Khoudli, L. Bresque, and L. Sanchez-Palencia, Critical behavior and fractality in shallow one-dimensional quasiperiodic potentials, Phys. Rev. Lett. {\bf 123}, 070405 (2019). 
\bibitem{Ganeshan2015} S. Ganeshan, J. H. Pixley, and S. Das Sarma, Nearest neighbor tight binding models with an exact mobility edge in one dimension, Phys. Rev. Lett. {\bf 114}, 146601 (2015).
\bibitem{Wang1} Y. Wang, X. Xia, L. Zhang, H. Yao, S. Chen, J. You, Q. Zhou, and X.-J. Liu, One dimensional quasiperiodic mosaic lattice with exact mobility edges, Phys. Rev. Lett. {\bf 125}, 196604 (2020).
\bibitem{XCZhou} X.-C. Zhou, Y. Wang, T.-F. J. Poon, Q. Zhou, and X.-J. Liu, Exact new mobility edges between critical and localized states, Phys. Rev. Lett. {\bf 131}, 176401 (2023).
\bibitem{Wang2022} Y. Wang, L. Zhang, W. Sun, T.-F. J. Poon, and X.-J. Liu, Quantum phase with coexisting localized, extended, and critical zones, Phys. Rev. B {\bf 106}, L140203 (2022).
\bibitem{XDeng2019} X. Deng, S. Ray, S. Sinha, G. Shlyapnikov, and L. Santos, One-dimensional quasicrystals with power-law hopping, Phys. Rev. Lett. {\bf 123}, 025301 (2019).
\bibitem{Ribeiro} M. Gon\c{c}alves, B. Amorim, E. V. Castro, and P. Ribeiro, Hidden dualities in 1D quasiperiodic lattice models, SciPost Phys. {\bf 13}, 046 (2022); M. Gon\c{c}alves, B. Amorim, E. V. Castro, and P. Ribeiro, Renormalization-Group Theory of 1D quasiperiodic lattice models with commensurate approximants, Phys. Rev. B {\bf 108}, L100201 (2023); M. Gon\c{c}alves, B. Amorim, E. V. Castro, and P. Ribeiro, Critical phase dualities in 1D exactly-solvable quasiperiodic models, Phys. Rev. Lett. {\bf 131}, 186303 (2023). 
\bibitem{Roati2008} G. Roati, C. D'Errico, L. Fallani, M. Fattori, C. Fort, M. Zaccanti, G. Modugno, M. Modugno, and M. Inguscio, Anderson localization of a non-interacting Bose-Einstein condensate, Nature (London) {\bf 453}, 895 (2008).
\bibitem{Bloch4} H. P. L\"{u}schen, S. Scherg, T. Kohlert, M. Schreiber, P. Bordia, X. Li, S. D. Sarma, and I. Bloch, Single-particle mobility edge in a one-dimensional quasiperiodic optical lattice, Phys. Rev. Lett. {\bf 120}, 160404 (2018);T. Kohlert, S. Scherg, X. Li, H. P. L\"{u}schen, S. D. Sarma, I. Bloch, and M. Aidelsburger, Observation of many-body localization in a one-dimensional system with single-particle mobility edge, Phys. Rev. Lett. {\bf 122}, 170403 (2019).
\bibitem{An2018} F. A. An, E. J. Meier, and B. Gadway, Engineering a flux-dependent mobility edge in disordered zigzag chains, Phys. Rev. X {\bf 8}, 031045 (2018); F. A. An, K. Padavi\'{c}, E. J. Meier, S. Hegde, S. Ganeshan, J. H. Pixley, S. Vishveshwara, and B. Gadway, Observation of tunable mobility edges in generalized Aubry-Andr\'{e} lattices, Phys. Rev. Lett. {\bf 126}, 040603 (2021).
\bibitem{JiasT} Y. Wang, J.-H. Zhang, Y. Li, J. Wu, W. Liu, F. Mei, Y. Hu, L. Xiao, J. Ma, C. Chin, and S. Jia, Observation of Interaction-Induced Mobility Edge in an Atomic Aubry-Andr\'{e} Wire, Phys. Rev. Lett. {\bf 129}, 103401 (2022).

\bibitem{TXiao2021} T. Xiao, D. Xie, Z. Dong, T. Chen, W. Yi, and B. Yan, Observation of topological phase with critical localization in a quasi-periodic lattice, Science Bulletin {\bf 66}, 2175
(2021).
\bibitem{Weld} T. Shimasaki, M. Prichard, H. E. Kondakci, J. Pagett, Y. Bai, P. Dotti, A. Cao, T.-C. Lu, T. Grover, and
D. M. Weld, Anomalous localization and multifractality in a kicked quasicrystal, arXiv:2203.09442.
\bibitem{HengFan} H. Li, Y.-Y. Wang, Y.-H. Shi, K. Huang, X. Song, G.-H. Liang, Z.-Y. Mei, B. Zhou, H. Zhang, J.-C. Zhang, et al., Observation of critical phase transition in a generalized Aubry-Andr\'{e}-Harper model with superconducting circuits,
npj Quantum Information 9, 40 (2023).
\bibitem{WangYC2021} Y. Wang, L. Zhang, S. Niu, D. Yu, X.-J. Liu, Realization and detection of non-ergodic critical phases in optical Raman lattice, Phys. Rev. Lett. {\bf 125}, 073204 (2020).
\bibitem{Kohmoto1990} Y. Hatsugai and M. Kohmoto, Energy spectrum and the quantum Hall effect on the square lattice with next-nearest-neighbor hopping, Phys. Rev. B {\bf 42}, 8282 (1990); J. H. Han, D. J. Thouless, H. Hiramoto, and M. Kohmoto, Critical and bicritical properties of Harper's equation with next-nearest-neighbor coupling, Phys. Rev. B {\bf 50}, 11365 (1994).
\bibitem{LandiRMP}  G. T. Landi, D. Poletti, and G. Schaller, Nonequilibrium boundary-driven quantum systems: Models, methods, and properties, Rev. Mod. Phys. {\bf 94}, 045006 (2022).
\bibitem{Saha} M. Saha, S. K. Maiti, and A. Purkayastha, Anomalous transport through algebraically localized states in one dimension, Phys. Rev. B {\bf 100}, 174201 (2019); A. Purkayastha, A. Dhar, and M. Kulkarni, Nonequilibrium phase diagram of a one-dimensional quasiperiodic system with a single-particle mobility edge,  Phys. Rev. B {\bf 96}, 180204(R) (2017); M. Saha, B. P. Venkatesh, and B. K. Agarwalla, Quantum transport in quasiperiodic lattice systems in the presence of B\"{u}ttiker probes, Phys. Rev. B {\bf 105}, 224204 (2022).
\bibitem{Dwiputra} D. Dwiputra and F. P. Zen, Environment-assisted quantum transport and mobility edges, Phys. Rev. A {\bf 104}, 022205 (2021).
\bibitem{Lacerda} A. M. Lacerda, J. Goold, and G. T. Landi, Dephasing enhanced transport in boundary-driven quasiperiodic chains, Phys. Rev. B {\bf 104}, 174203 (2021); V. Balachandran, S. R. Clark, J. Goold, and D. Poletti, Energy Current Rectification and Mobility Edges, Phys. Rev. Lett. {\bf 123}, 020603 (2019); C. Chiaracane, M. T. Mitchison, A. Purkayastha,
G. Haack, and J. Goold, Quasiperiodic quantum heat engines with a mobility edge, Phys. Rev. Research {\bf 2}, 013093 (2020).
\bibitem{Jeffrey} T.-F. J. Poon, Y. Wan, Y. Wang, and X.-J. Liu, Anomalous quantum transport in 2D asymptotic quasiperiodic system, arXiv:2312.04349.
\bibitem{IBlochRMP} D. A. Abanin, E. Altman, I. Bloch, and M. Serbyn, Colloquium: Many-body localization, thermalization, and
entanglement, Rev. Mod. Phys. {\bf 91}, 021001 (2019).
\bibitem{BAA} D. M. Basko, I. L. Aleiner, and B. L. Altshuler, Metal-insulator transition in a weakly interacting many-electron system with localized single-particle states, Ann. Phys. {\bf 321}, 1126 (2006).
\bibitem{Schreiber2015} M. Schreiber, S. S. Hodgman, P. Bordia, H. P. L\"{u}schen, M. H. Fischer, R. Vosk, E. Altman, U. Schneider, and I. Bloch, Observation of many-body localization of interacting fermions in a quasirandom optical lattice, Science {\bf 349}, 842 (2015).
\bibitem{SJAhn} S. J. Ahn, et al, Dirac electrons in a dodecagonal graphene quasicrystal Science {\bf 361}, 782 (2018).
\bibitem{BHuang2019} B. Huang and W. V. Liu, Moir\'{e} localization in two-dimensional
quasiperiodic systems, Phys. Rev. B {\bf 100}, 144202 (2019).
\bibitem{DMao2021} D. Mao and T. Senthil, Quasiperiodicity, band topology, and
moir\'{e} graphene, Phys. Rev. B {\bf 103}, 115110 (2021).
\bibitem{Castro2022} M. Gon\c{c}alves, H. Z. Olyaei, B. Amorim, R. Mondaini, P. Ribeiro, and E. V. Castro, Incommensurability-induced sub-ballistic narrow-band-states in twisted bilayer graphen, 2D Mater. {\bf 9}, 011001 (2022).
\bibitem{AUri2023} A. Uri, et al, Superconductivity and strong interactions in a tunable moir\'{e} quasicrystal, Nature {\bf 620}, 762 (2023). 
\bibitem{LiangFu} N. Paul, P. J. D. Crowley, and L. Fu, Dimensional reduction from magnetic field in Moir\'{e} superlattice, arXiv:2311.09323.
\bibitem{Shklovskii1993} B. I. Shklovskii, B. Shapiro, B. R. Sears, P. Lambrianides, and H. B. Shore, Statistics of spectra of disordered systems near the metal-insulator transition, Phys. Rev. B {\bf 47}, 11487 (1993).
\bibitem{Mirlin2000} A. D. Mirlin, Statistics of energy levels and eigenfunctions in disordered systems, Phys. Rep. 326, 259 (2000).
\bibitem{Molcanov1981} S. A. Mol\v{c}anov, The Local Structure of the Spectrum of the One-Dimensional Schr\"{o}dinger Operator, Commun. Math. Phys. {\bf 78}, 429 (1981).

\bibitem{XLi2016} X. Li, J. H. Pixley, D.-L. Deng, S. Ganeshan, and S. Das Sarma, Quantum nonergodicity and fermion localization in a system with a single-particle mobility edge, Phys. Rev. B {\bf 93}, 184204 (2016).
\bibitem{YLiu2020} Y. Liu, X.-P. Jiang, J. Cao, and S. Chen, Non-Hermitian mobility edges in one-dimensional quasicrystals with parity-time symmetry, Phys. Rev. B {\bf 101}, 174205 (2020).
\bibitem{YWang2023} Y. Wang, L. Zhang, Y. Wan, Y. He, and Y. Wang, Two-dimensional vertex-decorated Lieb lattice with exact mobility edges and robust flat bands, Phys. Rev. B {\bf 107}, L140201 (2023).
\bibitem{Xianlong} S. Cheng, R. Asgari, and G. Xianlong, From topological phase to transverse Anderson localization in a two-dimensional quasiperiodic system, Phys. Rev. B {\bf 108}, 024204 (2023).
\bibitem{Ray2016} S. Ray, M. Pandey, A. Ghosh and S. Sinha, Localization of weakly interacting Bose gas in quasiperiodic potential, New J. Phys. {\bf 18}, 013013 (2016).
\bibitem{JHPixley} J. H. Pixley, J. H. Wilson, D. A. Huse, and S. Gopalakrishnan, Weyl Semimetal to Metal Phase Transitions Driven by Quasiperiodic Potentials, Phys. Rev. Lett. {\bf 120}, 207604 (2018).
\bibitem{Schiffer2021} S. Schiffer, X.-J. Liu, H. Hu, and J. Wang, Anderson localization transition in a robust PT-symmetric phase of a generalized Aubry-Andr\'{e} model, Phys. Rev. A {\bf 103}, L011302 (2021).
\bibitem{Ray2018} S. Ray, A. Ghosh, and S. Sinha, Drive-induced delocalization in the Aubry-Andr\'{e} model, Phys. Rev. E {\bf 97}, 010101(R) (2018).
\bibitem{Khemani2017} V. Khemani, D. N. Sheng, and D. A. Huse, Two Universality Classes for the Many-Body Localization Transition, Phys. Rev. Lett. {\bf 119}, 075702 (2017).
\bibitem{Roushan2017} P. Roushan, C. Neill, J. Tangpanitanon, V.M. Bastidas, A. Megrant, R. Barends, Y. Chen, Z. Chen, B. Chiaro, A. Dunsworth, A. Fowler, B. Foxen, M. Giustina, E. Jeffrey, J. Kelly, E. Lucero, J. Mutus, M. Neeley, C. Quintana, D. Sank, A. Vainsencher, J. Wenner, T. White, H. Neven, D. G. Angelakis, J. Martinis, Spectroscopic signatures of localization with interacting photons in superconducting qubits, Science {\bf 358}, 1175 (2017).
\bibitem{XiaopengLi} X. Li, S. Ganeshan, J. H. Pixley, and S. D. Sarma, Many-Body Localization and Quantum Nonergodicity in a Model with a Single-Particle Mobility Edge, Phys. Rev. Lett. {\bf 115}, 186601 (2015).
\bibitem{Modak2015} R. Modak and S. Mukerjee, Many-Body Localization in the Presence of a Single-Particle Mobility Edge,
Phys. Rev. Lett. {\bf 115}, 230401 (2015).
\bibitem{Machida1986} K. Machida and M. Fujita, Quantum energy spectra and one-dimensional quasiperiodic systems, Phys. Rev. B {\bf 34}, 7367 (1986); M. Fujita and K. Machida, Spectral Properties of One-Dimensional Quasi-Crystalline and Incommensurate Systems, J. Phys. Soc. Jpn. {\bf 56}, 1470 (1987).
\bibitem{Vu2022} D. Vu, K. Huang, X. Li, and S. Das Sarma, Fermionic Many-Body Localization for Random and Quasiperiodic Systems in the Presence of Short- and Long-Range Interactions, Phys. Rev. Lett. {\bf 128}, 146601 (2022).
\bibitem{SNEvangelou} S. N. Evangelou and E. N. Economou, Spectral density correlations and eigenfunction fluctuations in one-dimensional quasi-periodic systems, J. Phys.: Condens. Matter {\bf 3} 5499 (1991).
\bibitem{Roy2019} N. Roy and A. Sharma, Study of counterintuitive transport properties in the Aubry-Andr\'{e}-Harper model via entanglement entropy and persistent current, Phys. Rev. B {\bf 100}, 195143 (2019).
\bibitem{Takada2004} Y. Takada, K. Ino, and M. Yamanaka, Statistics of spectra for critical quantum chaos in one-dimensional quasiperiodic systems, Phys. Rev. E {\bf 70}, 066203 (2004).
\bibitem{YWang2016} Y. Wang, Y. Wang, and S. Chen, Spectral statistics, finite-size scaling and multifractalanalysis of quasiperiodic chain with p-wave pairing, Eur. Phys. J. B {\bf 89}, 254 (2016).
\bibitem{Jitomirskaya} S. Jitomirskaya, Critical phenomena, arithmetic phase transitions, and universality: some recent
results on the almost Mathieu operator, {\em Current Developments in Mathematics} (2019); A. Avila and S. Jitomirskaya, In preparation.
\bibitem{AA} S. Aubry and G. Andr\'{e}, Analyticity breaking and Anderson localization in incommensurate lattices, Ann. Israel
Phys. Soc. {\bf 3}, 133 (1980).
\bibitem{DAHuse1} V. Oganesyan and D. A. Huse, Localization of interacting fermions at high temperature, Phys. Rev. B {\bf 75}, 155111 (2007).
\bibitem{DAHuse2} A. Pal and D. A. Huse, Many-body localization phase transition, Phys. Rev. B {\bf 82}, 174411 (2010).
\bibitem{Dyson1963} F. J. Dyson and M. L. Mehta, Statistical Theory of the Energy Levels of Complex Systems, J. Math. Phys. {\bf 4}, 701 (1963).
\bibitem{Guhr1998} T. Guhr, A. M\"{u}ller-Groeling, and H. A. Weidenm\"{u}ller, Random-matrix theories in quantum physics: common concepts, Phys. Rep. {\bf 299}, 189 (1998).
\bibitem{Bertrand2016} C. L. Bertrand and A. M. Garc\'{i}a-Garc\'{i}a, Anomalous Thouless energy and critical statistics on the metallic side of the many-body localization transition, Phys. Rev. B {\bf 94}, 144201 (2016).
\bibitem{WangMBC} Y. Wang, C. Cheng, X.-J. Liu, and D. Yu, Many-body critical phase: extended and nonthermal, Phys. Rev. Lett. {\bf 126}, 080602 (2021).
\bibitem{explain1} When $\theta_1$ satisfy $\theta_{1}\in[-2\pi/F_{N},0)$, one can easily find that there always exists an integer $n_0$ such that $n_{0}=(\theta+\pi-\theta_{1})/(\frac{2\pi}{F_{N}})$ holds true.
\bibitem{explaintheta} $\Theta(x)=0$ if $x\leq 0$ and $\Theta(x)=1$ if $x>0$. As shown in Fig.~\ref{03}(d), when the line of $\delta E$ intersects both the red and blue curves, $P(\delta E)$ is the sum calculated from both branches. However, when $\delta E$ only intersects the blue curve, for the red curve, $(b_{\kappa}L)^2-(\delta E)^2<0$. Therefore, $P(\delta E)$ only includes the blue branch in this case.

\bibitem{Kiczynski2022} M. Kiczynski, S. K. Gorman, H. Geng, M. B. Donnelly, Y. Chung, Y. He, J. G. Keizer, and M. Y. Simmons, Engineering topological states in atom-based semiconductor quantum dots, Nature (London) {\bf 606}, 694 (2022).
\end{thebibliography}
\end{document}